\documentclass[a4paper,fleqn,usenatbib]{mnras}
\usepackage[T1]{fontenc}
\usepackage{ae,aecompl}
\usepackage{comment}

\usepackage{graphicx}	
\usepackage{amsmath}	
\usepackage{amssymb}	
\usepackage{lineno}
\usepackage{epsfig}   
\usepackage{natbib}  
\usepackage{endnotes}
\usepackage{color}
\usepackage{fixfoot}
\usepackage{booktabs}

\title[Dynamic origin of Low-frequency QPOs]{Testing the dynamic origin of Quasi-periodic Oscillations in MAXI~J1535$-$571 and H 1743$-$322}

\author[Rawat et. al.]{Divya Rawat$^{1}$\thanks{E-mail: rawatdivya838@gmail.com (DR)},
Nazma Husain$^{2}$, and
Ranjeev Misra$^{1}$ \\
$^{1}$Inter-University Center for Astronomy and Astrophysics, Ganeshkhind, Pune 411007, India\\
$^{2}$Department of Physics, Jamia Millia Islamia, New Delhi 110025, India\\
}

\date{Accepted XXX. Received YYY; in original form ZZZ}

\pubyear{2023}

\begin{document}
\label{firstpage}
\pagerange{\pageref{firstpage}--\pageref{lastpage}}
\maketitle

\begin{abstract}
We investigate spectro-temporal properties for two black hole X-ray binary sources, MAXI~J1535$-$571 and H~1743$-$322,  during their hard and hard-intermediate states. For MAXI~J1535$-$571, we analyze {\sc{Swift/XRT}}, {\sc{NuSTAR}} and {\sc{NICER}} observations, specifically focusing on the occurrence of type-C Quasi-periodic Oscillations (QPOs). Regarding H~1743$-$322, we analyze multi-epoch observations of {\sc{NICER}} and AstroSat, identifying a type-C QPO with centroid frequency ranging from 0.1--0.6 Hz. In both sources, we fit the spectra with a relativistic truncated disk and a power law component. In MAXI J1535$-$571, we also observe an additional relativistically smeared iron line. Through temporal and spectral analysis, we estimate the QPO centroid frequency and spectral parameters, such as the accretion rate and inner disc radii. We test the origin of type-C QPOs as relativistic precession frequency, and dynamic frequency (i.e. the inverse of the sound crossing time $\frac{r}{c_s(r)}$). The dependence of QPO frequency on both the accretion rate and inner disc radii favours the QPO origin as dynamic frequency. We discuss the implications of these results.

\end{abstract}


\section{Introduction}
During an outburst, transient black hole X-ray binary (BHXB) sources make a transition to several canonical accretion states and, before returning to the quiescence state, trace a q-shape in the Hardness-Intensity Diagram \citep[HID;][]{ho01,fe04,be05}. These canonical accretion states are low hard state (LHS),  hard-intermediate state (HIMS), soft-intermediate state (SIMS), and high-soft state (HSS) \citep[][and references within]{be05,do07,be11}. The typical spectra of BHXBs constitute of disc black body component, the power-law component, which arises as a result of the comptonization of thermal photons and the reflection component, which is due to down-scattering of comptonized photons to the disc \citep[][and references within]{re06,do07}. In the LHS and HIMS, the spectra are dominated by emission from the power-law component, while in the SIMS and HSS, the thermal emission from the disc dominates \citep[for review, see][]{re06,do07,gi10}.\\

Transient BHXB sources also exhibit X-ray variability known as Quasi-periodic Oscillations (QPOs) on the time scale of mHz to kHz \citep[for review, see][]{in19}. The low-frequency QPOs (LFQPOs) are the ones with QPO frequency ranging from 0.1-20.0 Hz \citep[e.g.,][]{mo15}. Based on QPO's Q-factor and the presence of other harmonic components, the LFQPOs in these systems have been further classified into three categories, i.e., type-A, B and C (\citealt{wi99}, \citealt{ho01}, \citealt{re02}, \citealt{ca04}). The type-C QPO is the most commonly observed and appears in the hard or hard-intermediate state as narrow peaks in the Power Density Spectra \citep[PDS;][]{va85} accompanied with harmonic and broad Lorentzian noise components. Though the presence of type-C QPO  in the hard or hard-intermediate state of  BHXB systems is a ubiquitous phenomenon, there is still no consensus regarding the origin of these QPOs \citep{in19}.\\

One promising model that claims to explain the geometrical origin of  LFQPOs in compact objects is the relativistic precession model (RPM \citealt{st98,st99}).
In RPM, the QPO frequency is assumed to be 
nodal precession or Lense-Thirring  precession frequency ($\nu_{LT}$; \citealt{le18}). For a test particle of mass `M', orbiting in the tilted geodesics of characteristic radii `r', around a Kerr black hole of dimensionless spin parameter `a', the orbital,
and Lense-Thirring  precession frequencies are given by;
\begin{eqnarray}
    \nu_{\phi}=\pm \dfrac{c^{3}}{2 \pi G M M_\odot} \dfrac{1}{r^{3/2}\pm a}
\end{eqnarray}
\label{nu_phi}

\begin{eqnarray}
   \nu_{LT}=\nu_{\phi} [ 1-(1\mp \dfrac{4a}{r^{3/2}}+\dfrac{3 a^2}{r^2})^{1/2}]
 \label{nu_nod} 
\end{eqnarray}
where r is in units of gravitational radius, $R_g$. The LFQPO frequency has been identified as the Lense-Thirring  precession frequency for  BHXB sources GRO J1655$-$40, XTE J1550-564, MAXI J1820+070, and XTE J1859+226 \citep{mo14,mo14b,bh21,mo22}. Based on the detection of simultaneous LFQPO and HFQPOs, RPM allows for self-consistent measurements of the spin and mass of the black hole as well as estimating the radius at which the QPOs are produced (see \citealt{mo14,mo14b,mo22}). \\

There is another class of models  which assumes the dynamical mechanism as the origin of LFQPO instead of geometrical \citep{ca10,on11}. For the inner region of the standard thin relativistic accretion disc \citep{sh73,no73}, the dynamic frequency ($f_{dyn}$), which is inverse of the sound crossing time, is given by \cite{mi20} as;
\begin{eqnarray}
  \frac{f_{dyn}}{\dot M_{18}} & =  & N\hspace{0.1cm} 14030 \hspace{0.1cm}{\hbox{Hz}}\; (r/R_g)^{-2.5}\;(M/10 M_\odot)^{-2} \nonumber \\
  & & \times \;{\it A^1B^{-2}D^{-0.5}E^{-0.5}L}
  \label{fbyM}
\end{eqnarray}
Here $A$, $B$, $D$, $E$, $L $ are relativistic terms that depend on spin parameter `a' and  inner disk radii `r'.  `N' is the normalization factor \citep[please see][for more details]{mi20} and $\dot M_{18}$ is the accretion rate in terms of $10^{18}$ grams $s^{-1}$. \citet{mi20} and \citet{li21} have identified the LFQPO frequency as the relativistic dynamic frequency of a truncated accretion disc,  using AstroSat and Insight-HXMT observation of GRS 1915+105, respectively. \\

In this work, we investigate the origin of LFQPOs for two black hole transient sources, MAXI~J1535$-$571 and H~1743$-$322 using
two different LFQPO models; RPM, and dynamic origin. 
We briefly introduce the two BHXB sources in the following subsections.

\subsection{MAXI~J1535$-$571}

MAXI~J1535$-$571 (hereafter MAXI~J1535) is a transient low-mass BHXB source \citep{sc17,di17,ru17,ne17b} that was detected on 2$^{nd}$ Sep 2017 with MAXI/GSC \citep{ne17a} and SWIFT/BAT (\citealt{ke17}, \citealt{mr17}). Like other BHXB sources, this system also shows X-ray variability (\citealt{ge17a}, \citealt{me18}, \citealt{st18}, \citealt{hu18}, \citealt{bh19},\citealt{ne17b}, \citealt{zh22}, and \citealt{ra23}) and is found to trace a hysteresis loop in the HID \citep{ta18}. The source is located at a distance of 4--6 kpcs \citep{ca19,sr19} with a jet inclination angle $\le 45^\circ$ \citep{ru19}. X-ray spectral studies suggest that the system harbours a near-maximally spinning black hole (\citealt{ge17a}, \citealt{xu18}, \citealt{mi18}). \\

In its 2017 outburst starting from September 9-21, MAXI~J1535 was found to be in hard and hard-intermediate state (HIMS), and a type-C QPO was reported in the frequency range  0.2-9.0 Hz (\citealt{ge17a}, \citealt{me18}, \citealt{st18}, \citealt{hu18}, \citealt{bh19}, \citealt{ra23}). In the duration of HIMS, the photon index has been reported to be $\sim$2 with a low disk temperature of $\sim$0.3~keV \citep{ta18,sr19}. The source transitioned to a soft state and then to a soft-intermediate state with a  weak type-A/B LFQPO (\citealt{st18}, \citealt{se18}, \citealt{hu18}). During soft state, disk emission dominated the spectrum with disk temperature reaching $\sim$0.7~keV and inner radius extending up to ISCO \citep{ta18}. From $26^{th}$ September to $9^{th}$ October, the source transitions to the hard intermediate state with type-C QPO frequency ranging from 4.0-7.0 Hz (please see Figure 1 of \citealt{ra23}).
 
\subsection{H~1743-322}
 H~1743$-$322 (hereafter H1743) is an X-ray transient black hole binary source discovered and localized by \cite{ka77} and \cite{do77}. In 2003, the source was observed in its brightest and longest outburst and was extensively studied in multi-wavelengths \citep{pa03,ka06,ca09,co11}. H1743 was observed in several accretions states \citep{pa03} and showed all the three types of LFQPOs, i.e., type-A, B, and C \citep{ma03a,to03,ho05,ha22} during its 2003 outburst. After its 2003 complete outburst, the source has shown only two episodes of complete outburst in 2009 \citep{co11} and 2016 \citep{ch20} and several faint or failed outbursts, also termed as hard-only outbursts in 2004, 2005, 2008, 2010  (see Figure 1 and 2 of \citealt{co11}), 2012 \citep{sh14}, 2014 \citep{st16}, 2017 \citep{ji17}, and 2018 \citep{wi20,wa22}.\\

During the 2003 outburst of H1743, two-sided X-ray/radio jets were ejected from the source and, using symmetric kinematic models, \citet{st12} reported a distance of 8.5 kpcs and jet inclination angle of $75^\circ$. This reported value for the distance is consistent with the previously reported higher limit by \citet{co05}. Assuming the mass distribution given by \citet{oz10} for transient low mass X-ray binaries, \citet{st12} have constrained the spin of the black hole as -0.33 $\le$ a $\le$ 0.70 with a 90$\%$ confidence interval. With RXTE observations, the scaling technique between the spectral index and QPO frequency \citet{sh09} has estimated the dynamical mass of the black hole to be 11 $M_{\odot}$. In the hard state, the spectra of the source showed the presence of two distinct components, a multi-colour disc black body and hard power-law tail with photon index $\sim$2.2 \citep{ge03}. \\~\\
 
  Using {\sc{NICER}} observations, a variable type-C QPO for MAXI~J1535$-$571 (QPO$\sim$1.8-9.0 Hz; \citealt{ra23}) and H~1743$-$322 (QPO$\sim$0.1-0.6 Hz; \citealt{st21,wa22}) has been reported. We are analyzing the same set of {\sc{NICER}} observations of MAXI~J1535 as analyzed by \citet{ra23} for the study of the comptonizing medium through type-C QPOs with a complementary quasi-simultaneous {\sc{Swift/XRT}} and {\sc{NuSTAR}} observation. \citet{mi18} have analysed the {\sc{NICER}} observation (observation ID 1050360106) of the intermediate state and found clear hallmarks of relativistic disk reflection and emission components of the accretion disk and hard X-ray corona.
Additionally, the H1743 {\sc{NICER}} observations used in this work have been analysed by \citet{st21} and \citet{wa22} to study the spectral evolution of the source. However, this is the first time a spectro-temporal analysis of these two sources will be exploited to test the nature of type-C QPOs.  
The excellent time resolution of {\sc{NICER}} in the 0.2-10.0 keV energy band provides a unique opportunity to test the QPO models. As the 0.3-80.0 keV broad energy range and the good timing resolution of the LAXPC instrument onboard {\sc{AstroSat}} is best suited for the spectro-temporal study of the source, we have incorporated two {\sc{AstroSat}} H1743 observations for which type-C QPOs (QPO frequency$\sim$0.4-0.6 Hz) have been reported (\citealt{ch21} and \textcolor{blue}{Husain et al. 2023 under review}). Also, with low energy coverage of {\sc{SXT}} instrument, a better estimate of inner disc properties is possible as shown by \citet{mi20} (see right panel of  their Figure 4). \\

With timing analysis, we will first estimate the QPO centroid frequency. Next, we will fit the spectra with a relativistic truncated disc model and a relativistically smeared iron line model to estimate the inner disc radii and accretion rate. Finally, we will fit the RPM model (Equation \ref{nu_nod}) on the QPO frequency vs inner disc radii plot and dynamic frequency model (Equation \ref{fbyM}) on the QPO frequency/accretion rate vs inner disc radii plot. In Section 2, we describe the
observations and data analysis techniques, and in Section 3
we present the results of our analysis and fit the observational data points with Equations \ref{nu_nod} and \ref{fbyM}. Finally, we
discuss our findings in Section 4 and compare our results with microquasar source GRS 1915+105. In Section 5, we conclude our findings.

\begin{table}
 \caption{Observation log for MAXI~J1535$-$571; includes Instrument used, Observation ID, the start and end time of observation in M.J.D units, and exposure of observations.}
 \begin{center}
\scalebox{0.85}{%
\begin{tabular}{ccccc}
\hline
Instrument & ObsID & Tstart & Tstop & exposure \\ 
& & (M.J.D) & (M.J.D) &  (secs)\\ \hline 
\hline
{\sc{Swift/XRT}} & 00010264005 &58008.26 & 58008.29 & 2258 \\
\hline
{\sc{NuSTAR}} & 80302309002 & 58008.55 &58008.56 &881\\
 & & & &\\
 \hline
{\sc{NICER}} &1050360104 & 58008.46 & 58008.53 & 5310 \\
&1050360105 & 58008.99  & 58009.13 & 1509 \\
&1050360105 & 58009.16  & 58009.19 & 1554\\
&1050360105 & 58009.23  & 58009.30 & 791\\
&1050360105 & 58009.81  & 58009.94 &  2315\\
&1050360106 & 58010.00  & 58010.52 &  4838\\
&1050360107 & 58011.87  & 58011.94 & 1392\\
&1050360108 & 58012.19  & 58012.26 & 1219\\
&1050360108 & 58012.32  & 58012.58 &  1292\\
&1050360109 & 58013.22  & 58013.22 &  463\\
&1050360109 & 58013.28  & 58013.41 & 815\\
&1050360109 & 58013.48  & 58013.74 & 983\\
&1050360109 & 58013.99  & 58014.00 & 873\\
&1050360110 & 58014.05  & 58014.06 & 872\\
&1050360110 & 58014.82  & 58014.83 & 886\\
&1050360111 & 58015.28  & 58015.67 & 111\\
&1050360112 & 58016.24  & 58016.96 & 577\\
&1050360113 & 58017.01  & 58017.86 & 1254\\
&1130360103 & 58026.73  & 58026.81 &  87\\
&1130360104 & 58027.76  & 58027.78 & 1487\\
&1130360105 & 58028.72  & 58028.87 & 109\\
&1130360106 & 58029.75  & 58029.84 &  33\\
&1130360107 & 58030.71  & 58030.87 & 24\\
&1130360108 & 58031.36  & 58031.89 & 92\\
&1130360113 & 58036.50  & 58036.69 & 471\\
&1130360114 & 58037.03  & 58037.68 &  1272\\
\hline
\end{tabular}}
\end{center}
\label{table1}
\end{table}
\begin{table}
 \caption{Observation log of H~1743$-$322; includes Instrument used, Observation ID, the start and end time of observation in M.J.D units, and exposure of observations.}
 \begin{center}
\scalebox{0.85}{%
\begin{tabular}{ccccc}
\hline
Instrument & ObsID & Tstart & Tstop & exposure \\ 
& & (M.J.D) & (M.J.D) &  (secs)\\ \hline 
\hline
AstroSat & T01\textunderscore045T01\textunderscore9000000364  &  57456.41   &  57456.80    &   6246\\
& G07\textunderscore039T01\textunderscore9000001444 & 57973.32    &    57973.77    &   14490\\
\hline
{\sc{NICER}} & 1100300102 & 58367.01 & 58367.86 & 7781  \\
& 1100300104 & 58369.14 & 58369.92 & 1605  \\
& 1100300107 & 58372.75 & 58372.77 & 956  \\
& 1100300108 & 58373.06 & 58373.78 & 1548  \\
& 1100300115 & 58385.23 & 58385.82 & 2211  \\
\hline
\end{tabular}}
\end{center}
\label{table2}
\end{table}

\begin{figure}
\centering\includegraphics[scale=0.43,angle=0]{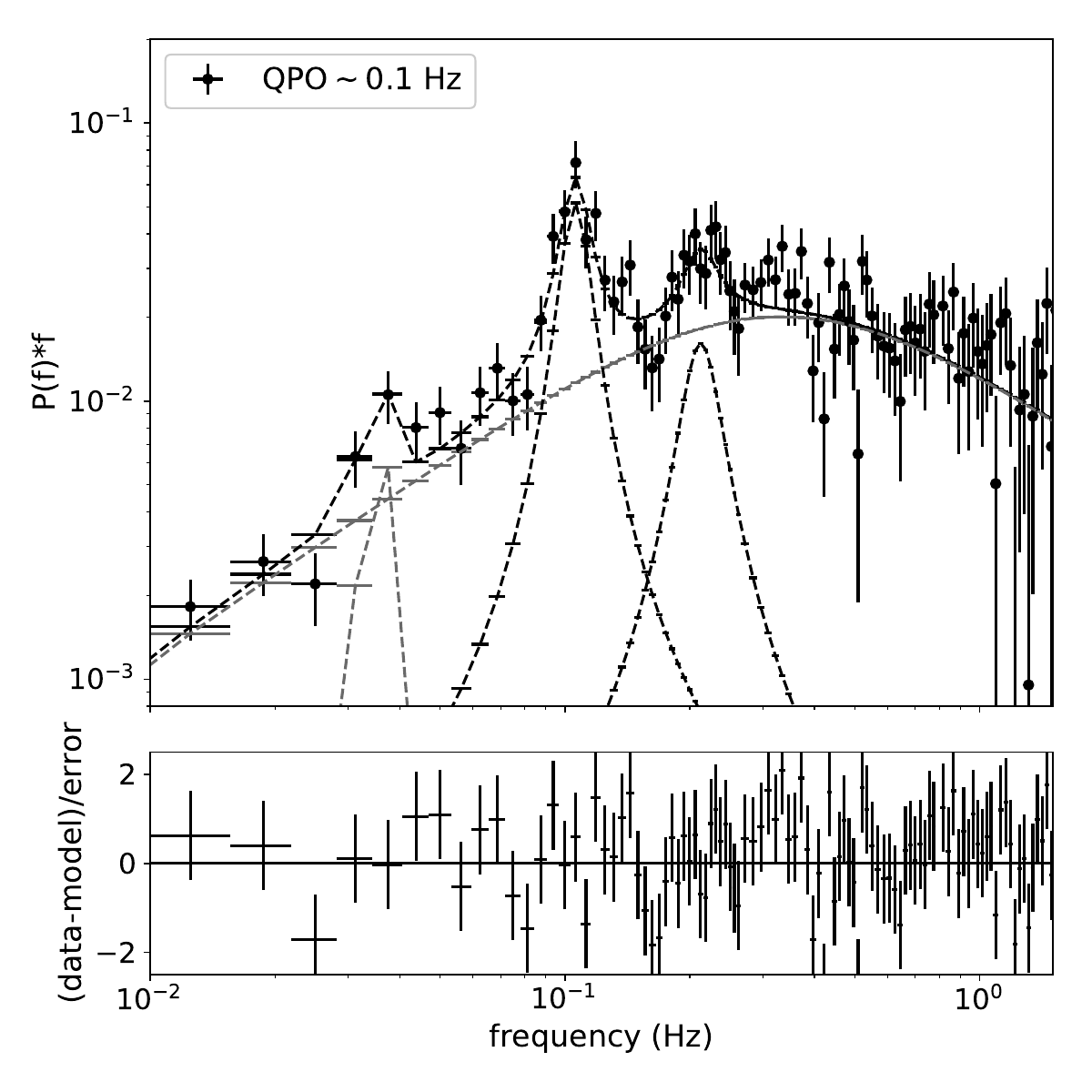}
\caption{The {\sc{NICER}} Power Density Spectra (PDS) of H~1743-322 fitted with 4 Lorentzians in 0.5--10.0 keV band is shown in the upper panel for observation ID 1100300102. The residuals of the fits for the PDS are shown in the bottom panel.}
\label{pds}
\end{figure}

\begin{figure*}
\centering\includegraphics[scale=0.395,angle=-90]{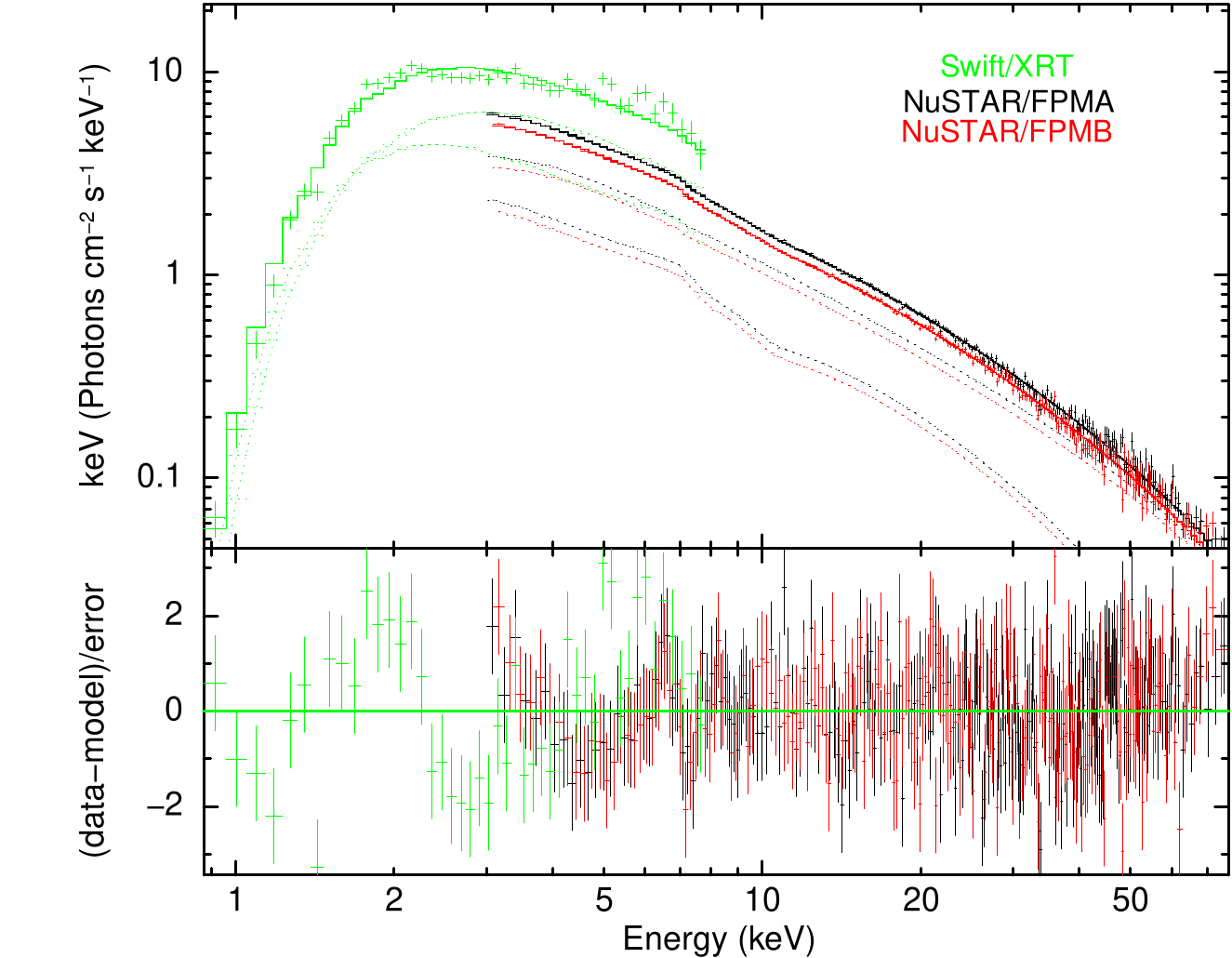}
\centering\includegraphics[scale=0.395,angle=-90]{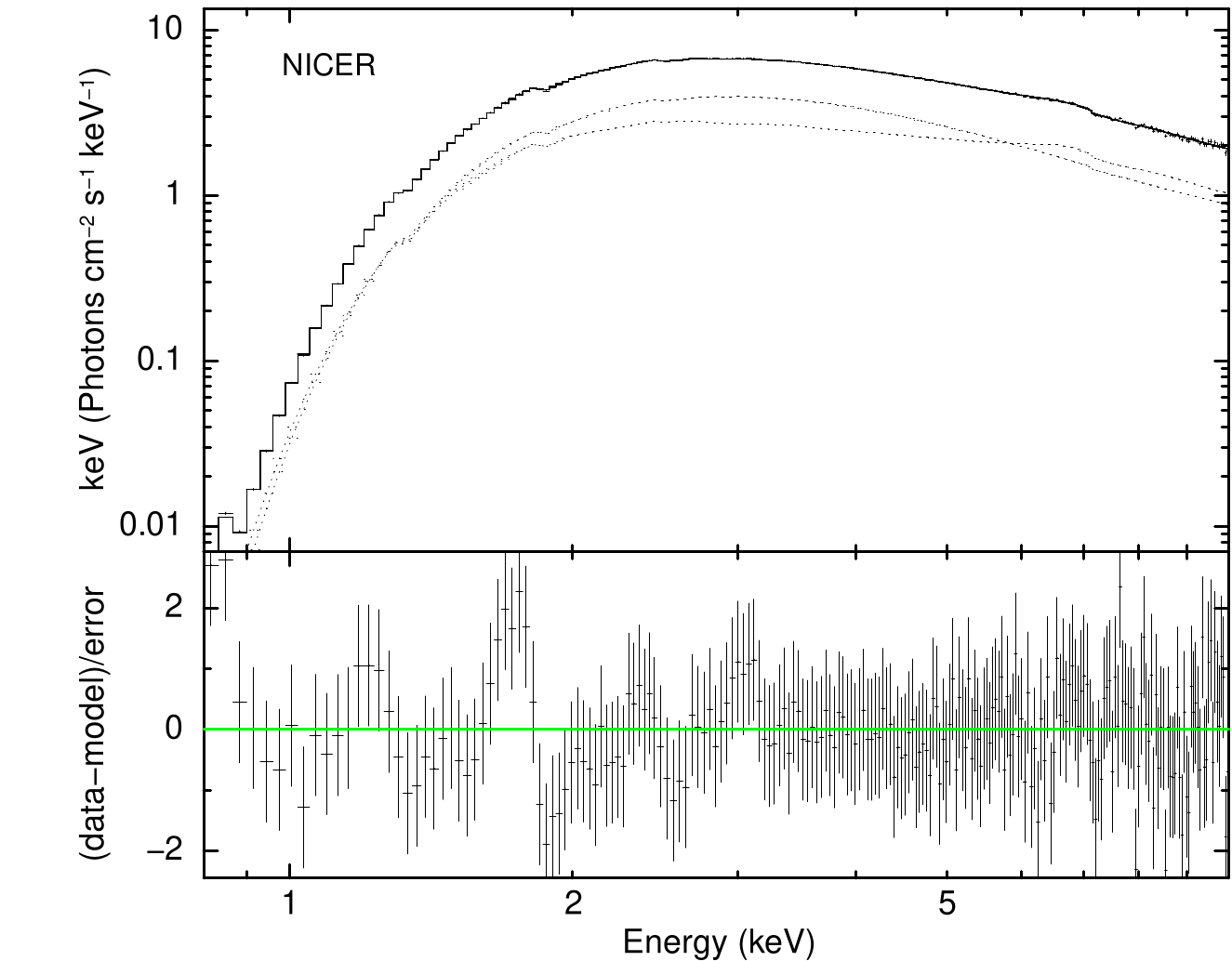}
\caption{The left panel shows the {\sc{Swift/XRT}} $\&$ {\sc{NuSTAR}} spectra of MAXI~J1535$-$571 simultaneously fitted with {\sc{TBabs(relxillCp+thcomp*kerrd}}) in 0.8--70.0 keV energy band. The right panel shows the {\sc{NICER}} spectra of MAXI J1535 fitted with the same model in the 0.8-79.0 keV band. The lower panels show the residual of the fit. The best-fit parameter values for both the spectra are shown in Table \ref{table3}.}
\label{Nustar_Spectra}
\end{figure*}
\section{Observation and  Data Reduction}
In this work, we have used {\sc{NICER}} extensive observations of MAXI~J1535 and H1743. We have also analyzed a nearly simultaneous {\sc{Swift/XRT}} and {\sc{NuSTAR}} observation for MAXI~J1535. The observation details for both sources are given in Table \ref{table1} and \ref{table2}.
\subsection{{\sc{Swift/XRT}}}
{\sc{Swift/XRT}} has observed the MAXI~J1535 in the window mode on 2017-09-12 06:16:57 with observation ID 00010264005 (details are given in Table \ref{table1}). {\sc{Swift/XRT}} \citep{burrow_2005} is an X-ray Imaging Telescope which operates in a narrow energy band of 0.2–10~keV with an effective area of $\sim$125 $cm^2$ at 1.5~keV.  It has a timing resolution of 2~ms. We have used Build Swift-XRT products\footnote{\label{note2}\url{https://www.swift.ac.uk/user_objects/}} to generate the spectra, ancillary response (arf) and  response (rmf; see \citealt{ev09}). By applying a threshold of >250 c/s \citep{mi07}, window mode effectively reduces pile-up for bright sources. To further correct for the effect, only grade 0 events were extracted for the source spectra. 

\subsection{{\sc{NuSTAR}}}
We have used the  Nuclear Spectroscopic Telescope Array ({\sc{NuSTAR}}) 2017-09-12 13:01:09 observation of MAXI~J1535 with observation ID 80302309002 (details are given in Table \ref{table1}). {\sc{NuSTAR}} is a high-energy X-ray mission that works in the 3.0-79.0 keV band. It consists of two focal plane module detectors, FPMA, and FPMB. In this work, we have used both module A and B to extract the level 2 products using {\sc{nupipeline}} from the level 1 fits file using the {\rm{saacalc=2, saamode = OPTIMIZED and tentacle =
no}}, flags. We have used ds9 to produce a source region file taking a circular source of region 3\arcmin. For background spectra, we have taken a circular region of 3\arcmin in the same frame away from the source. We then extract the spectra, arf and rmf using {\sc{nuproducts}} within {\sc{NuSTARDAS}}. \\

\subsection{{\sc{NICER}}}

In this work, we have used observations of MAXI~J1535 and H1743 obtained in September-October 2017 and September 2018, respectively, with the Neutron Star Interior Composition Explorer \citep[NICER][]{ge12}. The details of the observations ID's are given in Table \ref{table1} and \ref{table2}. {\sc{NICER}}'s XTI \citep[X-ray Timing Instrument][]{ge16} covers the 0.2-12.0 keV band and has an effective area of $>$2000 cm$^{2}$ at 1.5 keV. The energy and time resolution are 85 eV at 1 keV and 4 $\times 10^{-8}$ secs, respectively. We have used the {\sc{nicerl2}}\footnote{\label{note4}\url{https://heasarc.gsfc.nasa.gov/docs/nicer/analysis_threads/nicerl2/}} task to process each observation, which applies the standard calibration process and screening. We have produced the spectra and background files using the {\sc{NICER}} background estimator tool {\sc{3C 50\_RGv5}}\footnote{\label{note5}\url{https://heasarc.gsfc.nasa.gov/docs/nicer/tools/nicer_bkg_est_tools.html}}. We have used {\sc{Heasoft}} version 6.30 and CALDB version 20210707 to create the rmf and arf files.
For MAXI~J1535, the details of the spectra and Poisson noise subtracted PDS extraction are briefly given in \citet{ra23}. For H1743, we have fitted the PDS with 4 Lorentzians: one centered at 0 Hz, one for sub-harmonic component, one at QPO frequency and the last at a harmonic frequency of QPO. The fitted PDS for H1743 with reduced $\chi^{2}$ is shown in Figure \ref{pds}.

\begin{table*}
 \centering
\caption{Observation log for MAXI J1535$-$571 in the hard intermediate state, including selected parameters from the timing and spectral fits: QPO centroid frequency, nH, accretion rate, inner disc radii, power-law photon index, Flux in Line Emission, total unabsorbed Luminosity in 0.8-10.0 keV (Lum), reduced $\chi^{2}_{\nu}$ and degree of freedom, dof, for the  spectral fits. The errors are at 1$\sigma$. }
\begin{center}
\scalebox{0.75}{%
\begin{tabular}{cccccccccc}
\hline  
 Instrument  & QPO frequency &  $N_{\rm H}$   &  $\dot M_{18}$         & Inner Radius              & Fraction Scatter          & $\Gamma$                  & Flux in Line Emission & Lum    & $\chi^{2}_{\nu}$(dof) \\& (Hz) & $10^{22}$ cm$^{-2}$ & ($10^{18}$ gm s$^{-1}$) & (R$_{g}$) & & & ($10^{-2}$ photons cm$^{-2}$ s$^{-1}$)&  $10^{38}$ ergs cm$^{-2}$&
  \\ \hline 
  Swift+{\sc{NuSTAR}} & $2.15\pm{0.02}^{a}$ & $2.11\pm{0.04}$ & $0.55\pm{0.01}$ & $3.30^{+0.30}_{-0.09}$ & $\ge$0.97 &  $2.31\pm{0.01}$ & $16.9^{+0.01}_{-0.01}$ & $2.827\pm{0.015}$ & 516.5 (470)
\\~\\
  \hline
  {\sc{NICER}}  & $2.56\pm{0.00}$ & $2.28\pm{0.01}$ & $0.57\pm{0.06}$ & $2.3_{-2.3}^{+0.5}$ & $0.29_{-0.07}^{+0.11}$ & $1.91\pm{0.05}$ & $17.9_{-1.6}^{+2.4}$ & $2.855\pm{0.002}$  & $167.5 \;(216 \;)$ \\
& $2.74\pm{0.01}$ & $2.27\pm{0.01}$ & $0.59\pm{0.04}$ & $1.8_{-1.8}^{+0.8}$ & $0.08_{-0.04}^{+0.04}$ & $1.82\pm{0.02}$ & $22.2_{-2.4}^{+0.7}$ & $3.072\pm{0.003}$ & $142.2 \;(216 \;)$ \\
& $2.44\pm{0.01}$ & $2.28\pm{0.01}$ & $0.58\pm{0.06}$ & $2.5_{-2.5}^{+0.5}$ & $0.35_{-0.06}^{+0.16}$ & $1.93\pm{0.04}$ & $19.5_{-1.9}^{+1.6}$ & $3.018\pm{0.003}$ & $177.2 \;(216 \;)$ \\
& $2.32\pm{0.01}$ & $2.28\pm{0.01}$ & $0.56\pm{0.06}$ & $2.7_{-2.7}^{+0.4}$ & $0.50_{-0.14}^{+0.06}$ & $1.98\pm{0.04}$ & $18.2_{-1.1}^{+2.2}$ & $2.941\pm{0.003}$ &  $153.5 \;(216 \;)$ \\
& $1.83\pm{0.01}$ & $2.28\pm{0.01}$ & $0.42\pm{0.03}$ & $2.9_{-0.3}^{+0.3}$ & $0.84_{-0.06}^{+0.09}$ & $2.00\pm{0.03}$ & $19.5_{-0.8}^{+0.8}$ & $2.917\pm{0.003}$ & $162.0 \;(216 \;)$ \\
& $1.81\pm{0.00}$ & $2.27\pm{0.00}$ & $0.34\pm{0.01}$ & $1.2_{-1.2}^{+0.4}$ & $0.88_{-0.03}^{+0.03}$ & $2.00\pm{0.01}$ & $20.0_{-0.3}^{+0.5}$ & $2.790\pm{0.002}$ &  $156.5 \;(216 \;)$ \\
& $2.15\pm{0.01}$ & $2.23\pm{0.01}$ & $0.57\pm{0.02}$ & $2.3_{-0.7}^{+0.4}$ & $0.48_{-0.16}^{+0.08}$ & $1.92\pm{0.05}$ & $19.3_{-0.6}^{+2.5}$ & $3.143\pm{0.003}$ & $163.5 \;(216 \;)$ \\
& $2.41\pm{0.01}$ & $2.28\pm{0.01}$ & $0.62\pm{0.03}$ & $1.9_{-1.9}^{+0.6}$ & $0.17_{-0.04}^{+0.07}$ & $1.83\pm{0.03}$ & $25.6_{-1.6}^{+0.8}$ & $3.516\pm{0.003}$ & $132.2 \;(216 \;)$ \\
& $2.77\pm{0.01}$ & $2.28\pm{0.01}$ & $0.77\pm{0.11}$ & $2.5_{-2.5}^{+0.4}$ & $\leq0.14$ & $1.77\pm{0.06}$ & $26.5_{-2.6}^{+3.0}$ & $3.675\pm{0.003}$   & $151.1 \;(216 \;)$ \\
& $2.75\pm{0.02}$ & $2.28\pm{0.01}$ & $0.78\pm{0.05}$ & $2.2_{-0.3}^{+0.7}$ & $\leq0.13$ & $1.74\pm{0.04}$ & $29.6_{-5.4}^{+0.7}$ & $3.805\pm{0.004}$ & $163.7 \;(216 \;)$ \\
& $3.27\pm{0.02}$ & $2.29\pm{0.00}$ & $1.18\pm{0.02}$ & $3.2_{-0.1}^{+0.1}$ & $\leq0.02$ & $1.69\pm{0.01}$ & $28.2_{-0.1}^{+0.3}$ & $4.084\pm{0.004}$ & $154.9 \;(216 \;)$ \\
& $3.19\pm{0.03}$ & $2.28\pm{0.00}$ & $1.14\pm{0.01}$ & $3.1_{-0.1}^{+0.1}$ & $\leq0.02$ & $1.68\pm{0.01}$ & $29.6_{-1.1}^{+0.1}$ & $4.171\pm{0.004}$ & $142.8 \;(216 \;)$ \\
& $2.72\pm{0.01}$ & $2.27\pm{0.01}$ & $0.85\pm{0.03}$ & $2.5_{-0.1}^{+0.3}$ & $\leq0.06$ & $1.72\pm{0.05}$ & $31.2_{-2.8}^{+3.4}$ & $3.930\pm{0.004}$ & $129.4 \;(216 \;)$ \\
& $2.84\pm{0.01}$ & $2.28\pm{0.01}$ & $0.93\pm{0.04}$ & $2.7_{-0.1}^{+0.2}$ & $\leq0.05$ & $1.72\pm{0.02}$ & $30.8_{-2.5}^{+0.4}$ & $4.030\pm{0.004}$ & $149.5 \;(216 \;)$ \\
& $4.75\pm{0.01}$ & $2.32\pm{0.00}$ & $2.25\pm{0.03}$ & $4.3_{-0.1}^{+0.1}$ & $\geq 0.50$ & $1.56\pm{0.03}$ & $28.1_{-1.3}^{+0.9}$ & $4.983\pm{0.005}$ & $196.3 \;(216 \;)$ \\
& $9.01\pm{0.04}$ & $2.41\pm{0.01}$ & $3.83\pm{0.03}$ & $5.1_{-0.1}^{+0.1}$ & $\geq0.50$ & $1.20\pm{0.61}$ & $44.2_{-4.3}^{+1.9}$ & $5.810\pm{0.008}$ & $276.0 \;(217 \;)$ \\
& $7.54\pm{0.05}$ & $2.38\pm{0.00}$ & $3.62\pm{0.03}$ & $5.1_{-0.1}^{+0.1}$ & $\leq0.01$ & $1.40\pm{0.02}$ & $27.8_{-1.4}^{+4.5}$ & $5.898\pm{0.006}$ & $308.5 \;(217 \;)$ \\
& $7.54\pm{0.06}$ & $2.39\pm{0.01}$ & $3.96\pm{0.09}$ & $5.3_{-0.1}^{+0.1}$ & $\leq0.02$ & $1.28\pm{0.09}$ & $28.9_{-18.0}^{+7.0}$ & $5.853\pm{0.005}$ & $285.7 \;(213 \;)$ \\
& $7.09\pm{0.03}$ & $2.36\pm{0.01}$ & $2.79\pm{0.05}$ & $5.1_{-0.1}^{+0.1}$ & $\geq 0.29$ & $1.48\pm{0.05}$ & $21.6_{-3.2}^{+3.1}$ & $4.696\pm{0.008}$ & $280.2 \;(216 \;)$ \\
& $5.42\pm{0.01}$ & $2.32\pm{0.00}$ & $2.06\pm{0.02}$ & $4.7_{-0.1}^{+0.1}$ & $--$ & $1.63\pm{0.01}$ & $20.7_{-0.6}^{+0.2}$ & $4.281\pm{0.004}$ & $220.5 \;(212 \;)$ \\
& $5.73\pm{0.01}$ & $2.35\pm{0.01}$ & $2.35\pm{0.01}$ & $4.9_{-0.1}^{+0.1}$ & $\leq0.02$ & $1.55\pm{0.05}$ & $20.1_{-3.3}^{+1.5}$ & $4.319\pm{0.007}$ & $202.3 \;(216 \;)$ \\
& $6.77\pm{0.02}$ & $2.37\pm{0.01}$ & $2.54\pm{0.05}$ & $5.0_{-0.1}^{+0.1}$ & $\geq0.50$ & $1.49\pm{0.10}$ & $16.9_{-2.9}^{+3.1}$ & $4.275\pm{0.011}$ & $197.3 \;(208 \;)$ \\
& $4.57\pm{0.01}$ & $2.34\pm{0.01}$ & $1.57\pm{0.05}$ & $4.5_{-0.1}^{+0.1}$ & $\leq0.01$ & $1.68\pm{0.04}$ & $20.9_{-8.7}^{+0.6}$ & $3.728\pm{0.011}$ & $213.8 \;(206 \;)$ \\
& $4.82\pm{0.01}$ & $2.34\pm{0.01}$ & $2.09\pm{0.03}$ & $4.9_{-0.1}^{+0.1}$ & $\leq0.02$ & $1.58\pm{0.05}$ & $18.0_{-1.7}^{+24.5}$ & $3.946\pm{0.007}$ & $211.4 \;(216 \;)$ \\
& $5.19\pm{0.03}$ & $2.29\pm{0.00}$ & $1.85\pm{0.01}$ & $4.9_{-0.1}^{+0.1}$ & $\leq0.01$ & $1.60\pm{0.03}$ & $13.7_{-0.5}^{+5.7}$ & $3.388\pm{0.005}$ & $217.3 \;(216 \;)$ \\
& $4.50\pm{0.01}$ & $2.31\pm{0.01}$ & $1.18\pm{0.02}$ & $4.3_{-0.1}^{+0.1}$ & $\geq0.50$ & $1.74\pm{0.01}$ & $16.7_{-0.1}^{+0.1}$ & $3.042\pm{0.003}$ & $199.7 \;(212 \;)$ \\
\hline
\end{tabular}}
\end{center}
      \small
    Notes:  $^a$QPO frequency value taken from Table 1 of \citet{me18} for {\rm{{\sc{Swift/XRT}}}} observation ID 00010264005.
\label{table3}
\end{table*}

\begin{figure*}
\centering\includegraphics[scale=0.395,angle=-90]{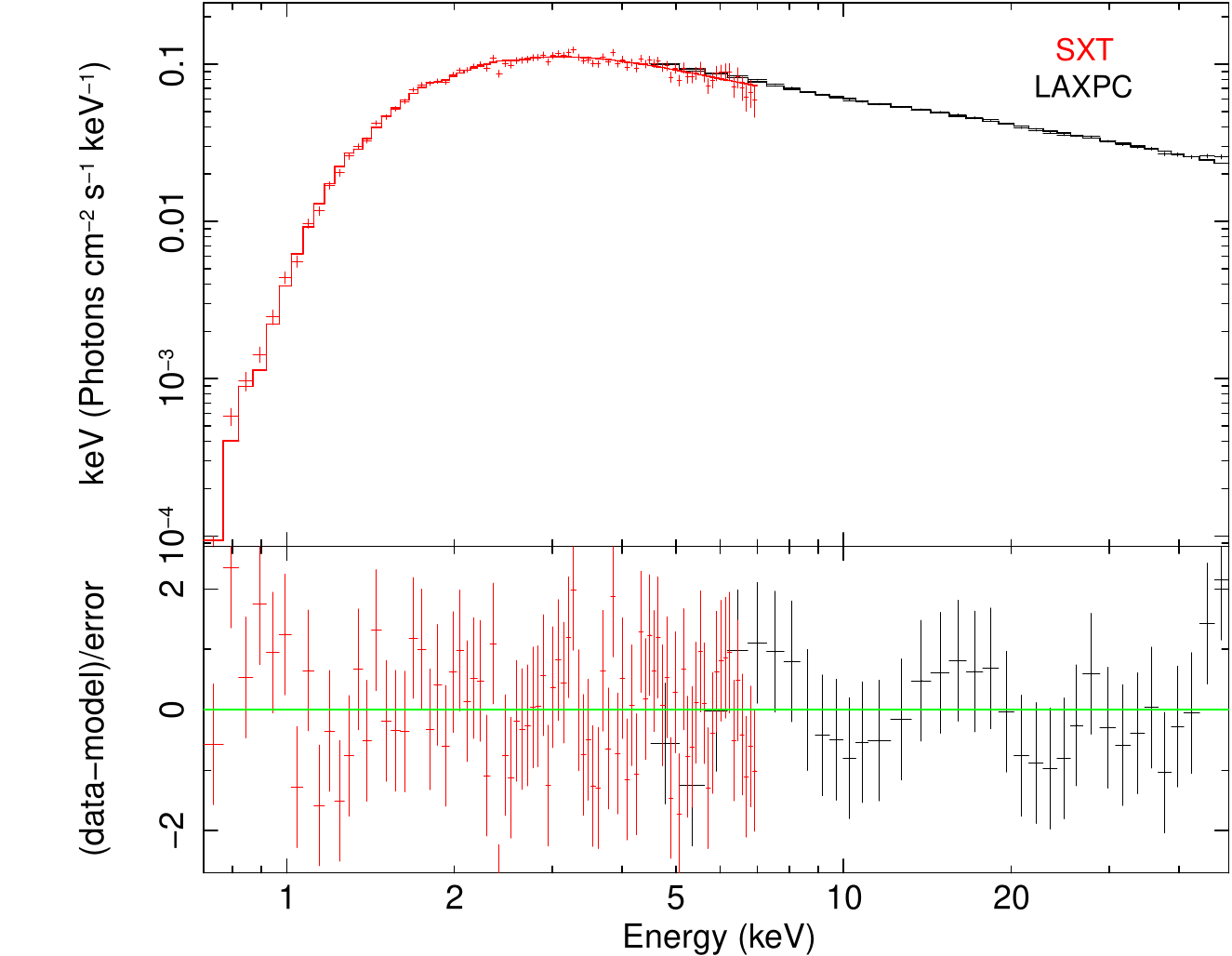}
\centering\includegraphics[scale=0.395,angle=-90]{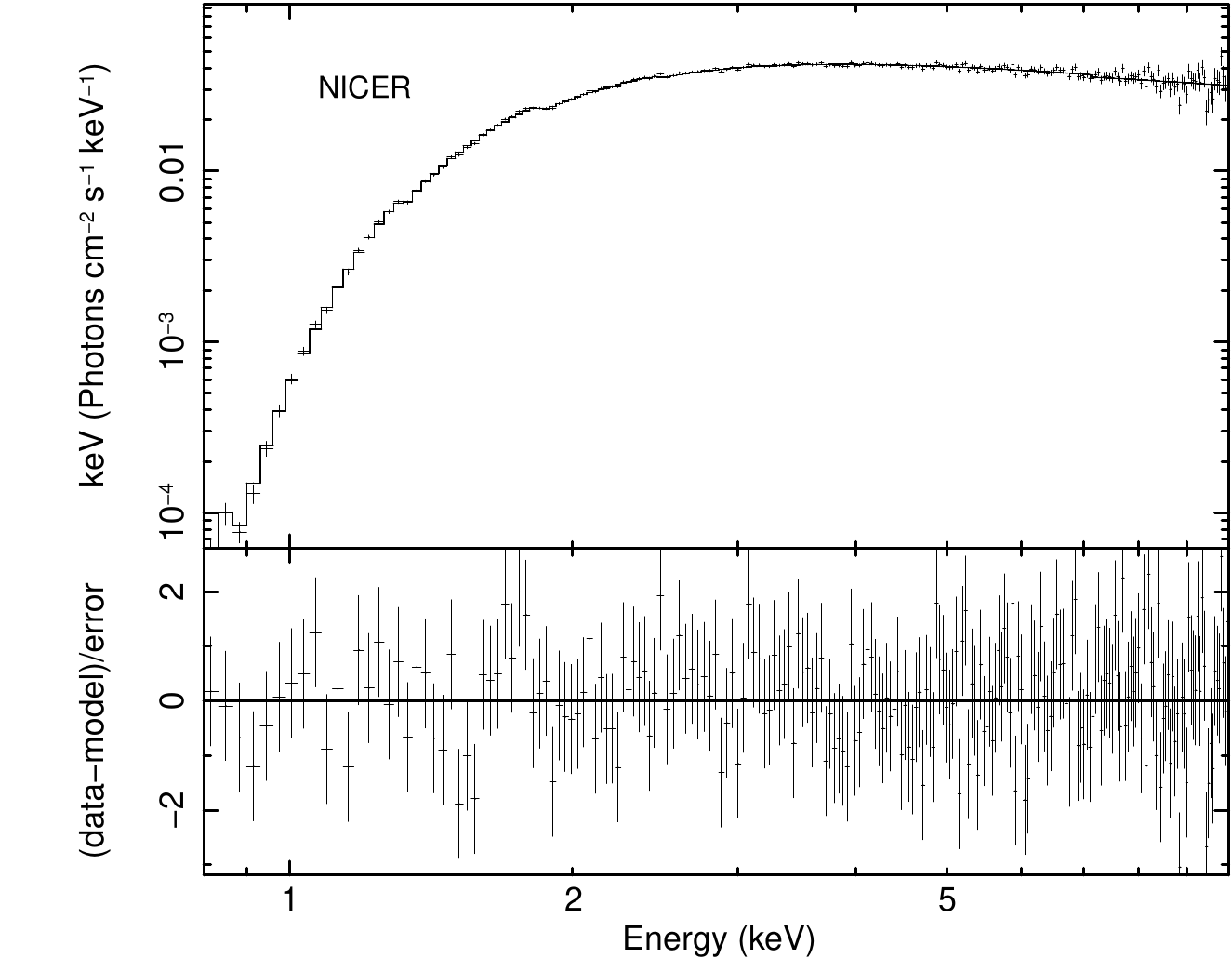}
\caption{The left panel shows the {\sc{AstroSat}} time-averaged spectra of H~1743-322 fitted with {\sc{Tbabs(thcomp*kerrd)}} model in the 0.7-50.0 keV energy band for observation ID T01$\_$045T01$\_$9000000364. The right panel shows the {\sc{NICER}} spectra of H~1743-322 fitted with the same model in the 0.8-10.0 keV energy band for observation ID 1100300102. The residuals of the fits for the respective spectra are shown in the bottom panels.}
\label{H1743_spectra_figure}
\end{figure*}

\begin{table*}
 \centering
\caption{Observation log for H~1743$-$322 in the hard state, including selected parameters from the timing and spectral fits: QPO centroid frequency, $N_{\rm H}$, accretion rate, inner disc radii, power-law photon index, total unabsorbed Luminosity in 0.8-10.0 keV (Lum), reduced $\chi^{2}_{\nu}$ and degree of freedom, dof, for the  spectral fits. The errors are at 1$\sigma$.}
\begin{center}
\scalebox{0.88}{%
\begin{tabular}{cccccccccc}
\hline  
 Instrument  &QPO frequency &  $N_{\rm H}$                      & Accretion Rate            & Inner Radius              & Fraction Scatter          & $\Gamma$ &    Lum   & $\chi^{2}_{\nu}$(dof) \\ & (Hz) & $10^{22}$ cm$^{-2}$ & ($10^{18}$ gm s$^{-1}$) & (R$_{g}$) & & & ($10^{37}$ ergs cm$^{-2}$)  & \\ \hline 
AstroSat &    $0.600\pm{0.002}$ & $2.18\pm{0.07}$ & $0.10\pm{0.01}$ & $2.9_{-0.4}^{+0.4}$ & $0.72\pm{0.03}$ & $1.64\pm{0.01}$ & $1.445\pm{0.006}$ & $103.6 \;(115 \;)$ \\
 & $0.440\pm{0.002}$ & $2.21\pm{0.07}$ & $0.08\pm{0.01}$ & $2.7_{-0.4}^{+0.5}$ & $0.77\pm{0.04}$ & $1.62\pm{0.01}$ & $1.307\pm{0.005}$ & $84.7 \;(114 \;)$ \\

 \hline
{\sc{NICER}} & $0.104\pm{0.002}$ & -- & $0.31\pm{0.04}$ & $26.5_{-2.3}^{+3.4}$ & $0.49\pm{0.03}$ & $1.51\pm{0.01}$ & $0.660\pm{0.001}$ & $221.9 \;(219 \;)$ \\
 & $0.151\pm{0.002}$ & -- & $0.68\pm{0.08}$ & $38.5_{-3.0}^{+4.4}$ & $0.45\pm{0.03}$ & $1.56\pm{0.01}$ & $0.997\pm{0.003}$ & $224.5 \;(218 \;)$ \\
 & $0.257\pm{0.002}$ & -- & $0.76\pm{0.08}$ & $30.2_{-3.5}^{+8.6}$ & $0.48\pm{0.02}$ & $1.56\pm{0.02}$ & $1.389\pm{0.005}$ & $233.8 \;(217 \;)$ \\
 & $0.258\pm{0.001}$ & -- & $0.94\pm{0.11}$ & $36.8_{-4.2}^{+4.9}$ & $0.46\pm{0.02}$ & $1.61\pm{0.01}$ & $1.380\pm{0.004}$ & $221.7 \;(218 \;)$ \\
 & $0.306\pm{0.007}$ & -- & $0.53\pm{0.03}$ & $23.9_{-1.5}^{+4.6}$ & $0.51\pm{0.01}$ & $1.59\pm{0.01}$ & $1.162\pm{0.003}$ & $202.3 \;(219 \;)$ \\
\hline
\end{tabular}}
\end{center}
\label{H1743_spectra_table}
\end{table*}

\subsection{AstroSat}
We have used AstroSat observations of the transient system H1743 as it went into an outburst on two separate events on 2016-03-09 and 2017-08-08 with observation IDs T01$\_$045T01$\_$9000000364 and G07$\_$039T01$\_$9000001444, respectively. AstroSat is India's first multi-wavelength satellite with five detectors working in X-rays, UV and visible range. AstroSat observations used in this work were conducted by the Soft X-ray Telescope (SXT; \citealt{si17}) and the Large Area X-ray Proportional Counter (LAXPC; \citealt{ya16a, an17}),  which works in the energy band of  0.3-8.0~keV and 3-80~keV, respectively. The reduction of Level-1 LAXPC data was performed with the LAXPC software package available on {\sc{AstroSat Science Support Cell (ASSC) website}}\footnote{\url{http://astrosat-ssc.iucaa.in/laxpcData}}.\\

The PDS in the 3.0-15~keV range is extracted using {powspec tool} of the XRONOS software package using the LAXPC data (for details, please see \textcolor{blue}{Husain et al. 2023 under review)}. For SXT, the spectrum is extracted from the Level-2 processed event file with the XSELECT (V2.4m) package of HEASOFT considering an encircled region of 15 arcmins on imaging tool ds9. We have used recent {\sc{ARF}}, RMF and background files provided by the {\sc{SXT Team}}\footnote{\label{tifr_webpage}\url{https://www.tifr.res.in/~astrosat_sxt/dataanalysis.html}}. The effective area file is corrected for the source region and vignetting effect using `SXTARFmodule' tool$^{\ref{tifr_webpage}}$. Further, the spectrum is combined with the corrected ARF, response and background files. More details on data reduction can be found in \textcolor{blue}{Husain et al. 2023 (under review)}.

\subsection{Spectral Analysis}
\subsubsection{MAXI J1535}
{\sc{Swift/XRT}} and {\sc{NuSTAR}} spectra of MAXI~J1535 show a contribution from a multi-temperature disc and power law emission components. We fit the multi-temperature disc with {\sc{kerrd}} model \citep{eb03}, which considers an optically thick accretion disc around a Kerr black hole. The power-law emission is taken care of by {\sc {thcomp}} \citep{zd20}, which is a convolution model that comptonizes the seed photons coming from the disc. We observe a relativistically smeared iron line at $\sim$6.7~keV, which we fit with {\sc{relxillCp}} model. {\sc{Tbabs}} takes care of the interstellar absorption. The time-averaged spectrum of MAXI~J1535 is fitted with {\sc{Tbabs(thcomp*kerrd+relxillCp)}} model as shown in left Figure \ref{Nustar_Spectra}. 

While fitting the spectra, we add 1 $\%$ systematic  and a constant, fixing the constant to `1' for {\sc{NuSTAR}} and free for {\sc{Swift/XRT}}.
The reported distance of the source with radio and X-ray observations is in the range of 4-6 kpc \citep{ca19,sr19}. So, we have fixed the distance parameter of {\sc{kerrd}} to 5.0 kpc. For black hole binary sources emitting at $\sim$10$\%$ Eddington rate, \citet{sh95} reported 
spectral hardening factor, $T_{col}$/$T_{eff}$ of 1.7. We have fixed the $T_{col}$/$T_{eff}$ of {\sc{kerrd}} to 1.7. Till now, there is no optical mass measurement available for the black hole, and X-ray spectral studies report its conflicting estimates \citep{sha19,sh19,sr19}. So, we have extracted the spectra considering a black hole of 10 solar masses. The spin parameter of {\sc{RelxillCp}} is fixed at the near-maximal value of 0.994 \citep{mi18}.
The inclination angle of {\sc{kerrd}} and {\sc{relxillCp}} is fixed to 45\textdegree \citep{ru19} while the mass accretion rate and inner disc radii are allowed to be free parameters. The time-averaged spectra fitted values for these two parameters with error bars are provided in Table \ref{table3}. For consistency, the inner disc radius of {\sc{kerrd}} is tied to that of {\sc{relxillCp}}. The outer disc radius and normalisation factor are fixed at 10$^5$ R$_g$ and 1, respectively. To ensure that the {\sc{relxillCp}} model exclusively outputs the reflection component, we have fixed the $refl_{frac}$ to -1. We have tied the power-law index, $\Gamma_{\tau}$, {\sc{thcomp}} to power law index, $\Gamma$ of {\sc{RelxillCp}}. The best-fit parameter values for power law index value
is given in Table~\ref{table3}. The best-fit gives $kT_e$= $34.5^{+0.7}_{-2.1}$ and logx$_i$=$3.61^{+0.02}_{-0.11}$ and constant for {\sc{Swift/XRT}} $\sim$1.53. The reduced $\chi^2$ of the spectral fit  and the best-fit parameters with a 1-sigma error are given in 
 Table \ref{table3}.\\

We then fit the {\sc{NICER}} (0.8--10.0 keV) spectra with {\sc{Tbabs(thcomp*kerrd+relxillCp)}} model as shown in the right panel of Figure \ref{Nustar_Spectra}. We have also added 1$\%$ systematic error, as suggested by the NICER team. We could not constrain the high-energy cutoff (which is parameterised by the electron temperature, $kT_e$) of {\sc{thcomp}} using {\sc{NICER}} data; therefore, we freeze it to the time-averaged spectral fitted value we got from {\sc{Swift}} and {\sc{NuSTAR}} spectra ($\sim$34.5 keV). The logx$_i$ values are also fixed to the best-fit spectral value ($\sim$3.6) from the {\sc{NuSTAR}}, and  {\sc{Swift/XRT}} spectral fitting. For the spectral fits, the reduced $\chi^2$ varies in the range 0.6--1.4. The best-fit parameters with the goodness of the fit for each segment are shown in Table \ref{table3}.
 
\subsubsection{H1743}
We have fitted the simultaneous SXT and LAXPC broadband spectra of H1743 with {\sc{constant*Tbabs(thcomp*kerrd)}} model in the 0.7--80.0 keV energy band as shown in the left panel of Figure \ref{H1743_spectra_figure}. For both epochs of AstroSat observation, no significant reflection component is observed in the spectra. While fitting the spectra, we have added 3$\%$ systematic \citep{an21} and fixed the constant factor to 1 for {\sc{LAXPC}} and kept if free for {\sc{SXT}}. The constant factor for {\sc{SXT}} ranges from 0.9--1.1 for the two epochs. We have added a gain for SXT and  have fixed the slope to 1. We fit the spectra for two epochs keeping the gain offset free, and finally fix it to the averaged value of the two observations, i.e., 0.002. {\rm{nH}} of {\sc{Tbabs}} is kept as a free parameter, and for both epochs, the best fit gives {\rm{nH}} of 2.2 (please see Table \ref{H1743_spectra_table}). We have kept the low energy power-law index, {\rm{$\Gamma_{\tau}$}} and {\rm{cov\_{frac}}} of {\sc{thcomp}} as free parameters. The high-energy cutoff of {\sc{thcomp}} is fixed to 30 keV following \citet{wa22}.
We have fixed the inclination angle and distance parameter of {\sc{kerrd}} to 75\textdegree and 8.5 kpc \citep{st12}, respectively. As the dynamical mass measurement using optical observations is not reported for this source, we have fixed the mass of {\sc{kerrd}} to 10 $M_\odot$.
The spectral hardening factor of {\sc{kerrd}} is again fixed to 1.7.
The mass accretion rate and inner disc radii of {\sc{kerrd}} are free parameters. The outer disc radius and normalisation factor are fixed at 10$^5$ R$_g$ and 1, respectively. The best-fit parameters for time-averaged spectra with their 1 $\sigma$ errors are given in Table \ref{H1743_spectra_table}.\\~\\
We have fitted the {\sc{NICER}} spectra of H1743 in the $0.8-10.0$  keV energy band with {\sc{Tbabs(simpl*kerrd)}} model as shown in the right panel of Figure \ref{H1743_spectra_figure}. We have fixed {\rm{nH}} to the time-averaged value of the best fit we got from AstroSat observation. We have kept inner disc radii, accretion rate, $\Gamma$, and fraction scatter as a free parameter, and their time-averaged best spectral fit values with $\chi^2$ of the fits are given in Table \ref{H1743_spectra_table}. The best fit gives a reduced $\chi^2$ varying in the range of 0.9-1.1.

\begin{figure*}
\centering\includegraphics[scale=0.55]{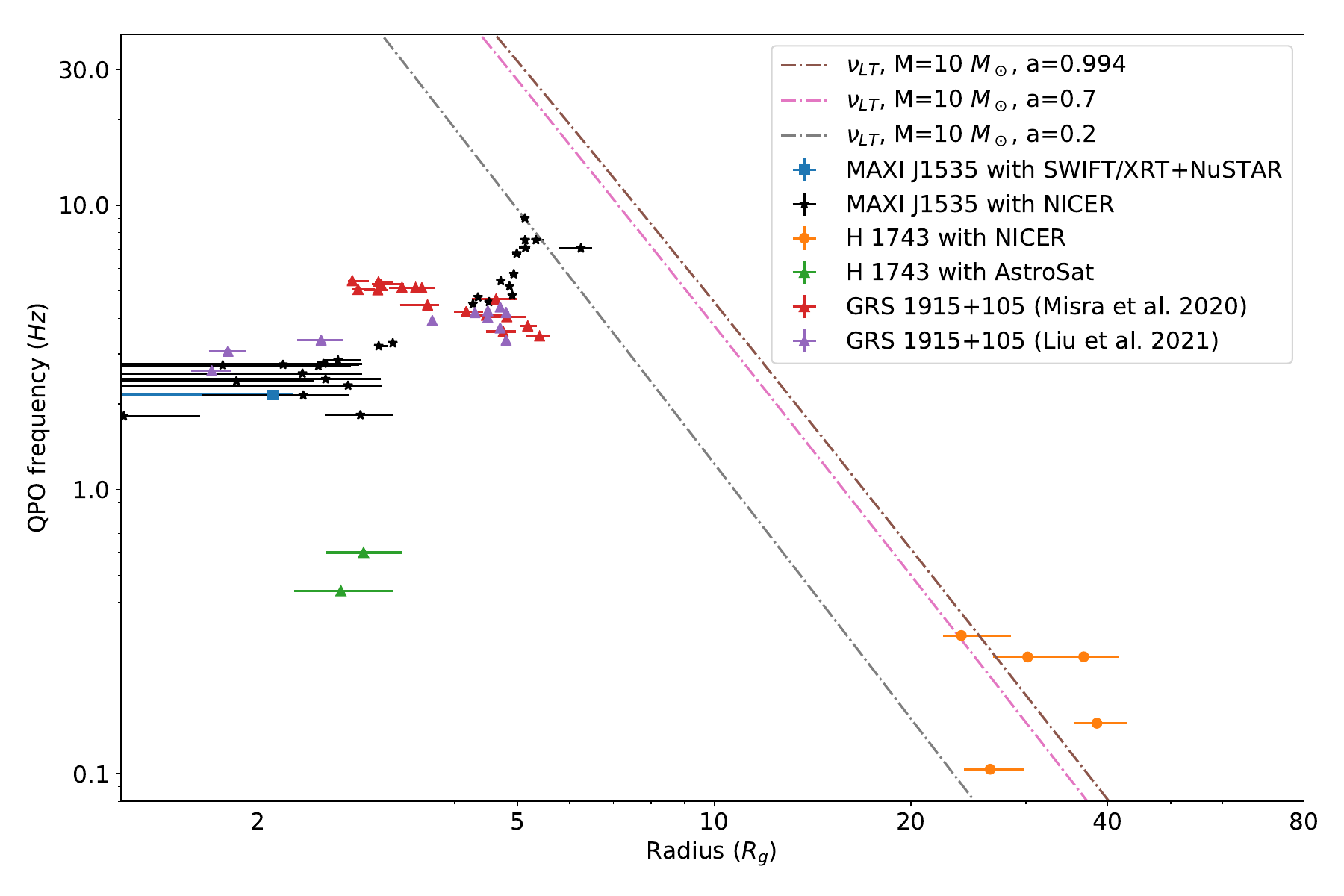}
\caption{The Figure shows the relationship between the QPO frequency and the inner disc radii for MAXI~J1535$-$571, H~1743$-$322 and GRS 1915+105. MAXI~J1535$-$571  data are represented by blue squares ({\sc{SWIFT/XRT+NuSTAR}}) and black data points ({\sc{NICER}}). {\sc{NICER}} and {\sc{AstroSat}} observations of H~1743$-$322 are shown as orange circles and green up-triangles. GRS 1915+105 data points from \citet{mi20} and \citet{li21} are red and purple up-triangles. The brown, magenta, and grey dash-dotted lines represent the Lense-Thirring frequency  (Equation \ref{nu_nod}) for a 10 10 $M \odot$ black hole with spin parameters of 0.994, 0.7, and 0.2, respectively.}
\label{lense_thirring}
 \small
    Note: We use the same colour scheme for observational data points in the next plots.
\end{figure*}

\begin{figure*}
\centering\includegraphics[scale=0.55]{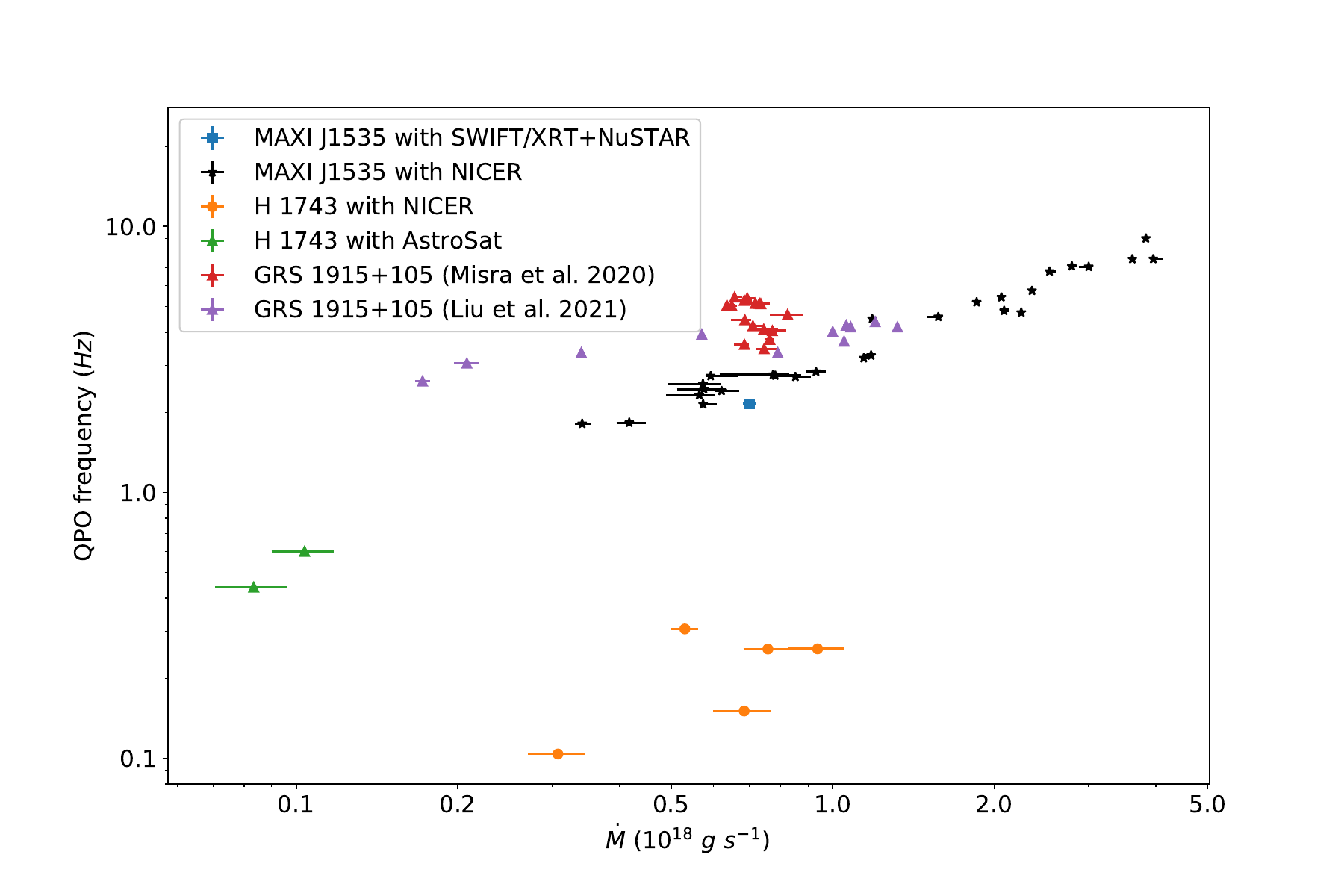}
\centering\includegraphics[scale=0.55]{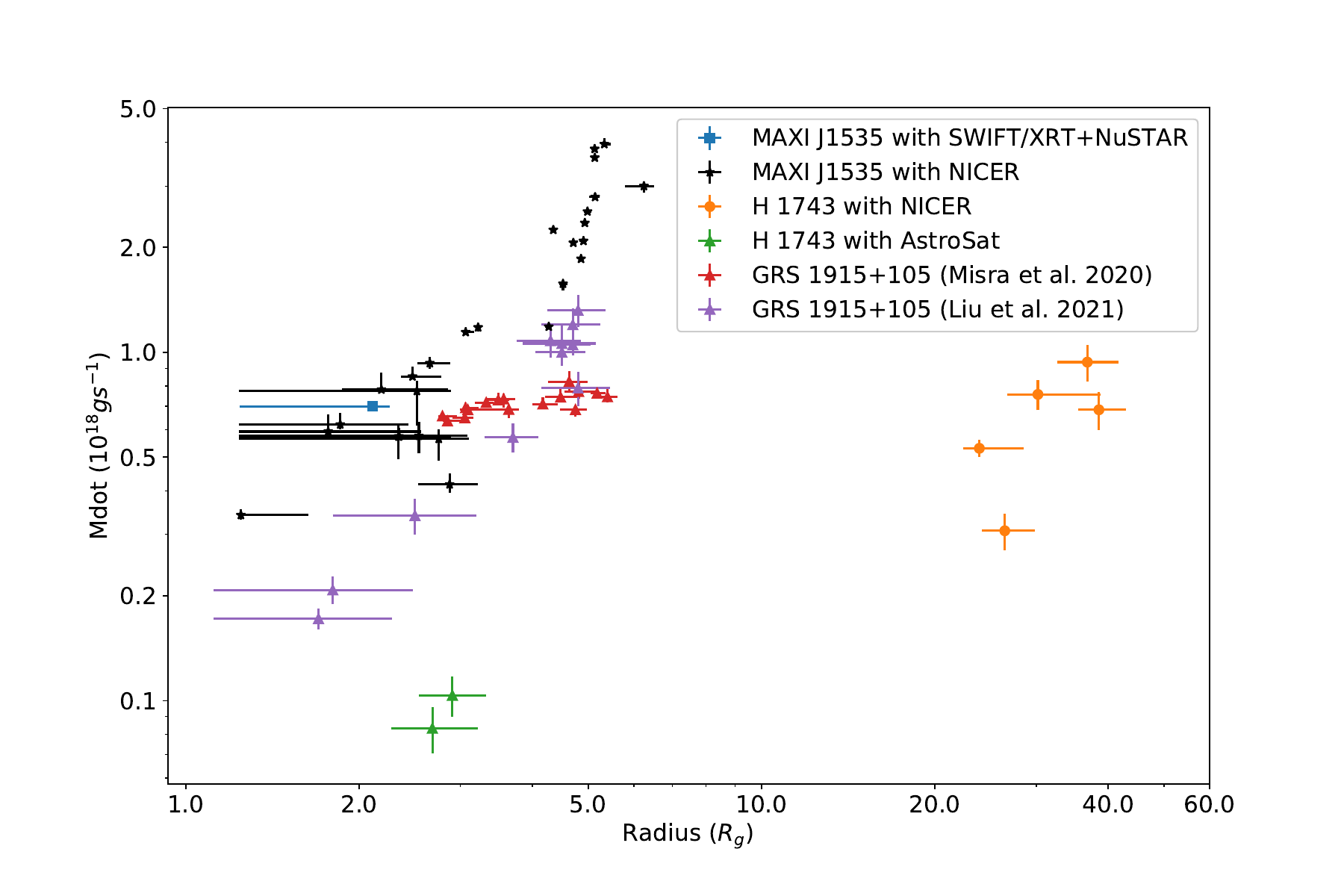}
\caption{The upper panel show the variation of QPO frequency with accretion rate, and the lower panel shows the accretion rate variation as a function of inner disc radii for MAXI~J1535$-$571 (blue and black coloured data points), H~1743$-$322 (orange and green and coloured data points) and GRS 1915+105 (red and purple coloured data points).}
\label{qpo_spec_par}
\end{figure*}
\begin{figure*}
\centering\includegraphics[scale=0.58,angle=0]{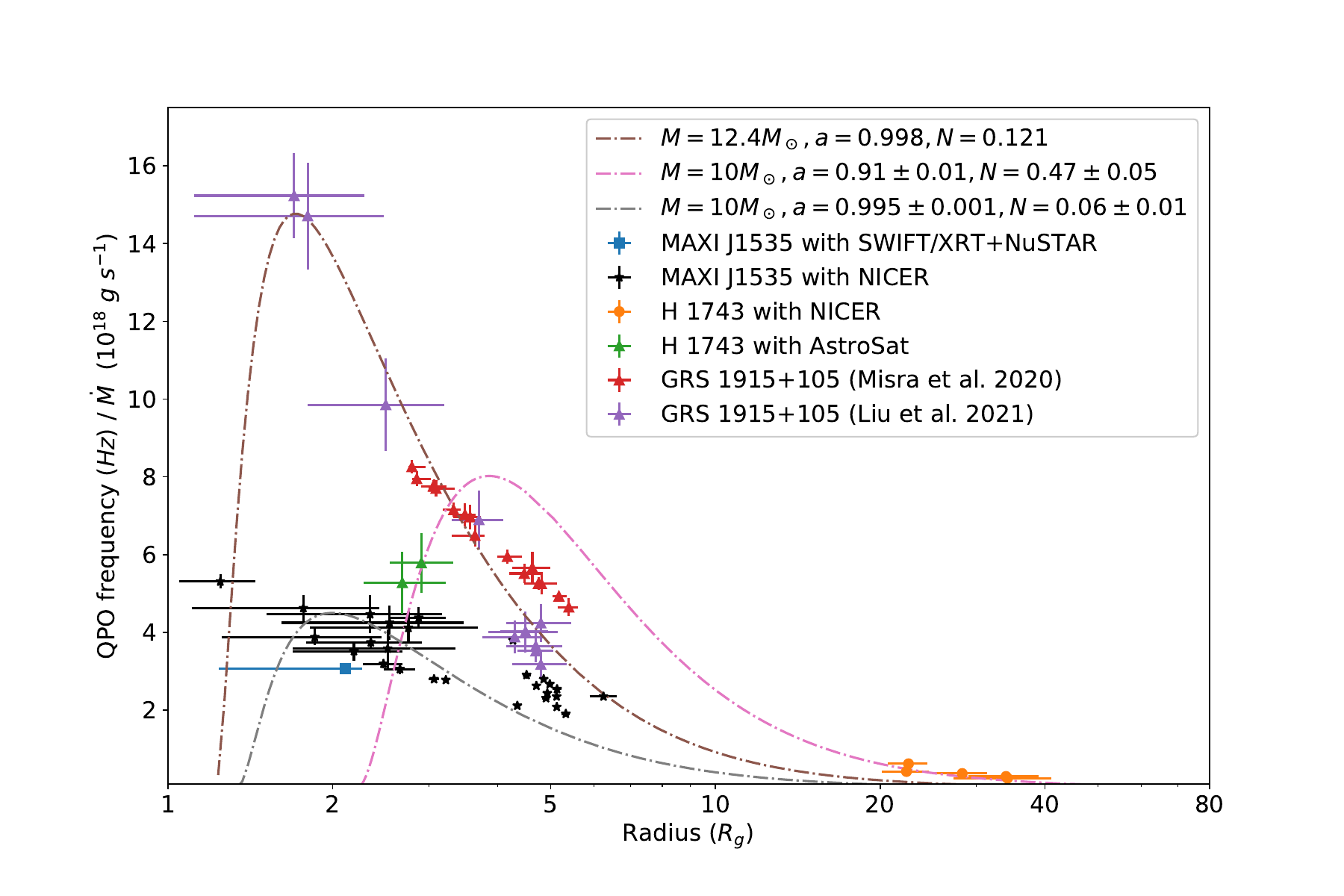}

\caption{The QPO frequency/accretion rate Versus inner disc radii plot for  MAXI~J1535$-$571, H~1743$-$322 and GRS 1915+105 are shown. The dash-dotted lines represent the theoretical curves for dynamic frequency/accretion rate at different values of the spin parameter, a, and normalization factor, N, given by Equation \ref{fbyM} for a 10 $M_\odot$ black hole.}
\label{test_dynamical}
\end{figure*}

\section{Results}
We have extracted the accretion rate and inner disc radii values from the spectral fit and the QPO centroid frequency value from the PDS fit. To study the  geometrical dependence of QPO frequency, we plot the QPO frequency as a function of inner disc radii as shown in Figure \ref{lense_thirring}. The black-coloured asterisk and blue-coloured square show {\sc{NICER}} and {\sc{Swift/XRT}}+{\sc{NuSTAR}} observation of MAXI J1535. The orange circular points and green triangle represent the  {\sc{NICER}} and AstroSat observation of H1743. For comparison, we over-plot the GRS 1915+105 AstroSat \citep{mi20} and HXMT \citep{li21} observations (shown in red and purple coloured triangles in Figure \ref{lense_thirring}). Using Equation \ref{nu_nod}, we have plotted curves for the nodal precession frequency as a function of inner disc radii for three different dimensionless spin parameter values, i.e., 0.994, 0.7, and 0.2. The nodal precession frequency curves are shown in brown, light pink and grey dash-dot lines for a 10 $M_{\odot}$ black hole. GRS 1915+105 and MAXI~J1535 are known to harbour a high-spin black hole; however, our results show that 
the data points for GRS 1915+105 and MAXI~J1535 don't coincide with the nodal precession frequency curve for a high-spin black hole. While in the case of H1743, the data points are scattered. The 2018's failed (or hard-only) state of H1743 (shown in orange triangles) aligns with the nodal precession frequency curve. In contrast, the data points from 2016's and 2017's observations fall much left to the nodal precession frequency in Figure \ref{lense_thirring}.\\~\\
Next, we plot the QPO frequency as a function of the accretion rate, as shown in the upper panel of Figure \ref{qpo_spec_par}. 
We used the same colour scheme and point type for data points as in Figure \ref{lense_thirring}. A broad correlation between QPO frequency and the accretion rate is observed for GRS 1915+105 and MAXI J1535. While again, for H1743, the data points are scattered. For MAXI J1535 and H 1743, the accretion rate varies in the range 0.04--0.28 $\dot M_{edd}$ and 0.007--0.062 $\dot M_{edd}$, respectively, assuming the radiative efficiency of 10\% for a 10 solar mass black hole.
Next, we checked the dependence of disc parameters, i.e., accretion rate and inner disc radii, with each other as shown in the lower panel of Figure \ref{qpo_spec_par}. There is a hint of a correlation between the
accretion rate and the inner disc radii for GRS~1915+105, MAXI J1535, and H1743.\\
To test the dynamic origin of the QPO, we plot the QPO frequency/$\dot M$ as a function of inner disc radii for GRS~1915+105, MAXI J1535, and H1743 as shown in Figure \ref{test_dynamical}. The dashed-dotted lines represent the theoretical curve for the dynamic frequency/$\dot M$, plotted as a function of inner disc radii as given by Equation \ref{fbyM}. 
The brown-coloured dash-dotted lines show the best-fit curve for GRS 1915+105 (taken form \citealt{mi20} and \citealt{li21}). We have kept `a' and `r' as free parameters in Equation \ref{fbyM} and used the mass of the black hole as 12.4 $M\odot$ for GRS 1915+105. The pink and grey coloured dashed-dotted lines show the best-fit  dynamic frequency curve for H1743 and MAXI~J1535, respectively. The black hole mass of H1743 and MAXI~J1535 is 10.0 $M\odot$. For MAXI~J1535, the data points are scattered, and the best fit gives dimensionless spin parameters, $a = 0.995 \pm 0.001$ and $N=0.06 \pm 0.01$.
For H1743, the orange-coloured data points are {\sc{NICER}} observations, and the green-coloured data points are AstroSat observations. In the hard-only state (orange-coloured data points), the inner disc is truncated at 22-34 R$_g$, so the {\sc{NICER}} data points are located at the lower end tail of the QPO frequency/$\dot M$ curve. However, in the hard-intermediate state, as observed with AstroSat (red-coloured data points), the inner disc is truncated at $\sim$3 R$_g$. The best fit gives $a=0.91\pm0.01$ and $N=0.47\pm 0.05$.\\
It should be noted that during the fitting process of the spectra, we kept the spectral hardening factor fixed at 1.7 for both sources. 
However, the spectral hardening factor could vary, leading to changes in the inner disk radii. To explore this possibility, we performed an analysis by extracting the inner disk radii with the spectral hardening factor fixed to two extreme values at 1.5 and 2.0. In the QPO frequency/accretion rate versus the inner disk radii plot, we observe a shift in the data points. Despite this shift in the data points, equation \ref{fbyM} still fits the data, yielding a spin parameter value nearly similar to that obtained by fixing the hardening factor to 1.7. Therefore, we conclude that the variation in the spectral hardening factor may not significantly impact our results.
\section{Discussion and Summary}
We have analysed the Sept-Oct 2017 observation of MAXI~J1535 using {\sc{NICER}} and {\sc{Swift/XRT}}+{\sc{NuSTAR}} instruments. We have also performed a spectral-temporal analysis for the H1743 using multi-epoch observations of its failed (2016, 2018) and complete outburst (2017) using AstroSat and {\sc{NICER}}. A variable type-C QPO in the 1.8-9.0 Hz frequency range was previously reported for the {\sc{NICER}} observation by \citet{ra23}; we have analysed the same observation set in which type-C QPO was present. Similarly, for H1743, a type-C QPO with QPO frequency 0.1-0.6 Hz was observed by \citet{wa22} and \citet{st21} for the {\sc{NICER}} observation set we have used in this work. We have fitted the spectra of MAXI~J1535 with a relativistic truncated disk model, a power-law component and a relativistically smeared iron lime emission component. 
\subsection{Geometrical origin of QPO frequency}
In the relativistic precession model (RPM), the LFQPO frequency is associated with the nodal precession frequency given by Equation \ref{nu_nod}. Our results show that the curve for the Lense-thirring frequency doesn't match the observational data points for MAXI~J1535, GRS 1915+105 and H1743. As shown in Figure \ref{lense_thirring}, the theoretical curve predicts a decrease in inner disc radii with an increase in the QPO frequency. However, for MAXI~J1535, a positive correlation between the QPO frequency and inner disc radii are found in this work. For, H1743, the data points are scattered.
Notice that the source is in a hard-intermediate state in its 2016 and 2017 outburst \citep{ch20}. While the 2018 outburst was a failed outburst as reported by \citet{wa22} and the source was in a hard state with a low accretion rate and truncated inner disc up to 22-34 $R_{g}$ (details are given in Table \ref{H1743_spectra_table}).\\

The differences in the slope of the Lense-thirring frequency and the observed data points for MAXI~J1535 and H1743 argues against the geometric origin of the QPO. Also, in the RPM model, the QPO frequency should be solely a function of inner disc radii and `a'. Instead, in this work, we have observed a correlation between the accretion rate and the QPO frequency. We discuss the implications of these results in the next subsection.
\subsection{QPO as a natural frequency in a disc-corona system}
For MAXI~J1535, we have observed a strong correlation between the QPO frequency and the accretion rate (as shown in Figure \ref{qpo_spec_par}). The over-plotted data points for GRS 1915+105 from \citet{mi20} and \citet{li21} also show a hint of correlation. Suppose the type-C LFQPOs originate from the radiative coupling between the hot comptonizing corona and the cold accretion disc \citep{ma22}. In that case, the QPO centroid frequency should be proportional to  $\sqrt{\dot{M}}$ (see Equation 15 of \citealt{ma22}), and the corona should extend upto several hundreds of $R_g$ ($\sim$600 R$_g$). For H1743, in the QPO frequency vs accretion rate plot, there is a change in slope between the AstroSat observations (green data points) and {\sc{NICER}} observations (orange data points in Figure \ref{qpo_spec_par}). For the AstroSat observations, the source was in the hard-intermediate state. While for {\sc{NICER}} observation, the source was in the hard-only state. Assuming the origin of QPO as natural frequency, this slope difference could be attributed to the two different coronal sizes.
In all the cases, the coronal size is significantly large,  extending a significant covering factor over the accretion disk. The large coronal sizes could result from the simple assumption of homogeneous spherical corona taken by \citet{ma22}. 

We have also reported a positive correlation between the accretion rate and inner disc radii for MAXI~J1535 and H1743. Using AstroSat observation of MAXI~J1535, \citet{ga22} have reported a correlation between the accretion rate and inner disc radii. This correlation was previously reported for the heartbeat state of GRS 1915+105 by \citet{ne12,ra22}. 
The spectral parameters, i.e., accretion rate and inner disc radii, should be noted, depending on the compact object's mass. For MAXI~J1535 and H1743, a dynamical mass measurement using optical observations is unavailable. So, the slope of the QPO frequency vs accretion rate plot for both sources also depends on the mass of the black hole, which is uncertain in this case. In the QPO models like RPM, where a geometric origin of QPO is proposed,  the QPO frequency should depend only on the truncated disc radii. The dependence of QPO frequency on inner disc radii and accretion rate provides strong evidence of the dynamic nature of the QPO. \\

\subsection{Dynamic origin of QPO frequency}
The theoretical curve for dynamic frequency overall fits observational data points for MAXI~J1535 for a dimensionless spin parameter of 0.994, previously reported by \citet{mi18}. Unlike GRS 1915+105, where the data points match well with the theoretical curve of dynamic frequency (see \citealt{mi20}), for MAXI~J1535, the data points are scattered. The scatter of data points for the MAXI~J1535 could be attributed to the limited energy range of the instrument plus the uncertainties in the calibration in the lower energy regime. As shown in the right panel of Figure 4 in \citet{mi20}, the low energy coverage of the SXT instrument plays a significant role in estimating inner disc properties like inner disc radii.
For H1743, in the hard-intermediate state, the disc is truncated to $\sim$3 R$_g$, and the location of data points suggests a higher spin value of the black hole than the previously reported values \citep{st12}. Notice that the value of spin reported by \citet{st12} highly depends on the black hole's mass, which is uncertain in this case. Plus, \citet{st12} has analysed RXTE observation of H1743, where a low energy coverage was missing, which plays an important role in estimating spin and inner disc parameters.\\

Notice that the spin parameter values reported using this method depend on parameters like the black hole's mass, inclination angle, distance to the source, uncertainties in ARFs and the colour factor. The dynamic frequency curve itself depends on the mass of the black hole. We show the dependence of black hole mass on the dynamic frequency curve for three black hole masses 10$M_{\odot}$, 20$M_{\odot}$, and 30$M_{\odot}$ in Appendix Figure \ref{mass_vary_dynm}, freezing the black hole spin to a=0.99 and N=0.12. This method requires low energy coverage for the spectral parameters estimation and the presence of significant QPOs. Plus, For  H 1743$-$322, we have tested the model in a narrow frequency range of QPO from 0.1--0.6 Hz. Our suggestion is to conduct further testing on the dynamic origin of QPOs in H 1743$-$322 and other black hole binary systems by performing simultaneous observations using different instruments, such as {\sc{AstroSat/LAXPC}} and {\sc{NICER}}. In addition, utilizing {\sc{NuSTAR}}'s high energy coverage will aid in resolving the spectral parameters.

\section*{Acknowledgements}
This work used data from the UK Swift Science Data Centre at the University of Leicester. This research has made use of the {\sc{NuSTAR}} data analysis software
(NuSTARDAS) jointly developed by the ASI science centre (ASDC, Italy) and the California Institute of Technology (Caltech, USA). The High Energy Astrophysics Science Archive Research Centre (HEASARC) and the Indian Space Science Data Centre (ISSDC) have provided the data for this research. DR acknowledges Yash Bhargava and Akash Garg for enlightening
discussions.

\section*{Data Availability}
The {\sc{NICER/XTI}}, {\sc{Swift/XRT}}, and {\sc{NuSTAR}} observations used in this work are available at \href{https://heasarc.gsfc.nasa.gov/db-perl//W3Browse/w3browse.pl}{HEASARC website}. The AstroSat data sets were derived from sources in the public domain: \href{https://webapps.issdc.gov.in/astro_archive/archive/Home.jsp}{AstroSat Archive}. The software and tools used to process the data are available at \href{http://astrosat-ssc.iucaa.in/?q=laxpcData}{LAXPC Software}, \href{http://astrosat-ssc.iucaa.in/?q=sxtData}{SXT Software}.


\def\aj{AJ} \def\actaa{Acta Astron.}  \def\araa{ARA\&A} \def\apj{ApJ}
\def\apjl{ApJ} \def\apjs{ApJS} \def\ao{Appl.~Opt.}  \def\apss{Ap\&SS}
\def\aap{A\&A} \def\aapr{A\&A~Rev.}  \def\aaps{A\&AS} \def\azh{AZh}
\def\baas{BAAS} \def\bac{Bull. astr. Inst. Czechosl.}
\def\caa{Chinese Astron. Astrophys.}  \def\cjaa{Chinese
  J. Astron. Astrophys.}  \def\icarus{Icarus} \def\jcap{J. Cosmology
  Astropart. Phys.}  \def\jrasc{JRASC} \def\mnras{MNRAS}
\def\memras{MmRAS} \def\na{New A} \def\nar{New A Rev.}
\def\pasa{PASA} \def\pra{Phys.~Rev.~A} \def\prb{Phys.~Rev.~B}
\def\prc{Phys.~Rev.~C} \def\prd{Phys.~Rev.~D} \def\pre{Phys.~Rev.~E}
\def\prl{Phys.~Rev.~Lett.}  \def\pasp{PASP} \def\pasj{PASJ}
\def\qjras{QJRAS} \def\rmxaa{Rev. Mexicana Astron. Astrofis.}
\def\skytel{S\&T} \def\solphys{Sol.~Phys.}  \def\sovast{Soviet~Ast.}
\def\ssr{Space~Sci.~Rev.}  \def\zap{ZAp} \def\nat{Nature}
\def\iaucirc{IAU~Circ.}  \def\aplett{Astrophys.~Lett.}
\def\apspr{Astrophys.~Space~Phys.~Res.}
\def\bain{Bull.~Astron.~Inst.~Netherlands}
\def\fcp{Fund.~Cosmic~Phys.}  \def\gca{Geochim.~Cosmochim.~Acta}
\def\grl{Geophys.~Res.~Lett.}  \def\jcp{J.~Chem.~Phys.}
\def\jgr{J.~Geophys.~Res.}
\def\jqsrt{J.~Quant.~Spec.~Radiat.~Transf.}
\def\memsai{Mem.~Soc.~Astron.~Italiana} \def\nphysa{Nucl.~Phys.~A}
\def\physrep{Phys.~Rep.}  \def\physscr{Phys.~Scr}
\def\planss{Planet.~Space~Sci.}  \def\procspie{Proc.~SPIE}
\let\astap=\aap \let\apjlett=\apjl \let\apjsupp=\apjs \let\applopt=\ao

\bibliographystyle{mnras}
\bibliography{manuscript} 




\appendix
\counterwithin{table}{section}

\section{ }
\begin{figure}
    \centering
    \includegraphics[scale=0.65,angle=0]{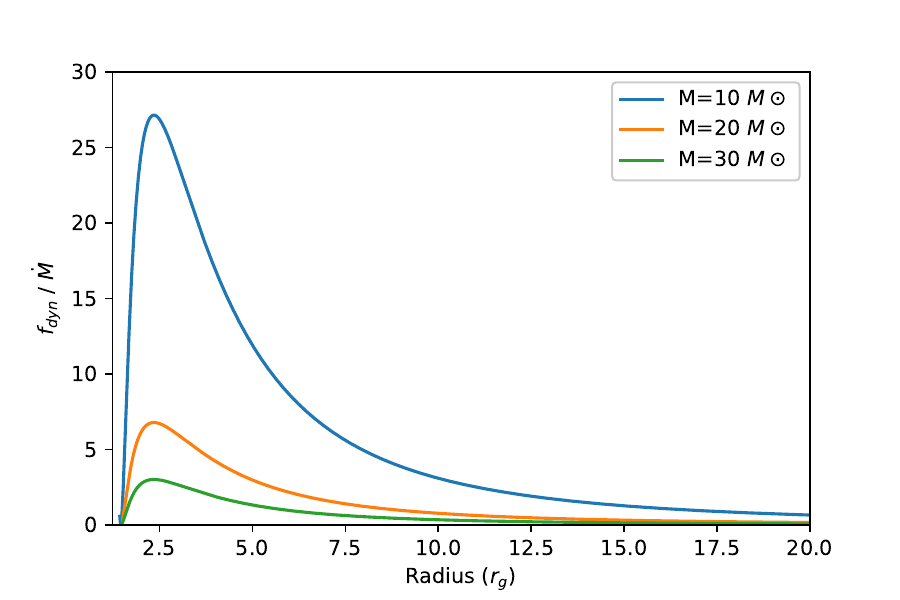}
        \caption{The dynamic frequency vs inner disk radii plot for three black hole masses, 10 $M_\odot$, 20 $M_\odot$, and 30 $M_\odot$ are shown in blue, orange and green colours solid lines. The spin parameter and normalization factor are fixed to a=0.99, N=0.12.}
    \label{mass_vary_dynm}
\end{figure}

In this Appendix, we show the dynamic frequency vs inner disk radii plot for three black hole masses, 10 $M_\odot$, 20 $M_\odot$, and 30 $M_\odot$ (Figure \ref{mass_vary_dynm}).
\counterwithin{figure}{section}

\label{lastpage}
\end{document}